\documentclass[conference]{IEEEtran}
\IEEEoverridecommandlockouts
\usepackage{cite}
\usepackage{amsmath,amssymb,amsfonts}
\usepackage{algorithmic}
\usepackage{graphicx}
\usepackage{booktabs}
\usepackage{url}
\usepackage[T1]{fontenc}
\usepackage{textcomp}
\usepackage{xcolor}
\def\BibTeX{{\rm B\kern-.05em{\sc i\kern-.025em b}\kern-.08em
    T\kern-.1667em\lower.7ex\hbox{E}\kern-.125emX}}
\begin{document}

\title{YouTube Clips and Broad but Shallow Attention to Candidates: Evidence from the 2024 Tokyo Gubernatorial Election\\
}

\author{\IEEEauthorblockN{1\textsuperscript{st} Yoshito Miwa}
\IEEEauthorblockA{\textit{Department of Innovation Science} \\
\textit{Institute of Science Tokyo}\\
Tokyo, Japan \\
miwa.y.bf00@m.isct.ac.jp}
\and
\IEEEauthorblockN{2\textsuperscript{nd} Kazutoshi Sasahara}
\IEEEauthorblockA{\textit{Department of Innovation Science} \\
\textit{Institute of Science Tokyo}\\
Tokyo, Japan \\
sasahara.k.6efa@m.isct.ac.jp}
}
\maketitle

\begin{abstract}
Video-sharing platforms and social media have gained influence in election campaigns, and third-party edited videos (called \textit{kirinuki} in Japanese; hereafter, “clip videos”) have attracted particular attention. In Japan, Shinji Ishimaru, previously unknown nationally, finished second in the 2024 Tokyo gubernatorial election, a result widely attributed to his popularity on YouTube. Focusing on this case, we analyzed 44,331 YouTube videos relating to the nine leading candidates, posted between July 5, 2020, and July 7, 2024, using regression analyses of cumulative views and engagement measured on June 25, 2025. Videos mentioning Ishimaru accumulated the most views, about 2.3 times those mentioning the second-ranked candidate. Under our main classification, clip videos not mentioning Ishimaru received about 1.8 times as many views as general videos, and non-clip videos mentioning Ishimaru received about 2.6 times as many views as those not mentioning him; the interaction between the two was not significant. Yet Ishimaru, the most watched candidate, did not draw the most active engagement. Over the full period, non-clip videos mentioning him received about 36\% fewer likes and 41\% fewer comments at the same number of views; this shortfall was concentrated in videos posted before the official campaign and reversed for likes among campaign-period videos. Clip videos not mentioning Ishimaru also received about 26\% fewer comments than general videos at the same number of views. These results suggest that, among videos posted before the campaign, attention to Ishimaru was broad but shallow in the sense measured here---many views with little engagement per view.
\end{abstract}


\section{Introduction}
 
Where people encounter political information is shifting from mass media to video platforms. In Japan, average daily internet use overtook television viewing on weekdays in 2020, and by 2024 it exceeded television viewing on both weekdays and holidays for all ages combined~\cite{soumu2025media}. Surveys of political communication likewise record a growing share of voters naming social media and video platforms among the sources they consult~\cite{nira2023}.
 
Among political content on video platforms, \emph{clip videos} (\textit{kirinuki})  have become increasingly prominent. In this study, a clip video is a video in which a third party reassembles footage that already exists online. Many are short edits that lift a striking passage out of a long livestream, a council session or a press conference and add subtitles or music. The format spread first in entertainment genres such as game commentary and has since moved into politics. Behind this lies YouTube's monetization scheme, under which
advertising revenue is shared according to view counts. Press reporting indicates that some operators of clip channels devoted to one candidate were motivated by revenue rather than political support, some of them earning several million yen a month~\cite{kyotonp2025kirinuki,nhk2025kirinuki}. Clip videos can thus be a mechanism by which third-party actors unaffiliated with the candidates generate political attention on a large scale, driven by financial incentives.
 
This phenomenon became visible in a Japanese election in the 2024 Tokyo gubernatorial race. Shinji Ishimaru, the former mayor of Akitakata (a city of fewer than 30,000 in Hiroshima Prefecture), was nationally unknown and received no organized party backing. Nevertheless, he took about 1.66 million votes to finish second behind the incumbent~\cite{tokyoelection2024}. The result was labeled the “Ishimaru phenomenon” and was widely attributed to his online popularity, centered on YouTube~\cite{mainichi2024ishimaru,go2senkyo2024}. About half of his supporters named YouTube among the sources they consulted before voting~\cite{yoneshige2024}. Quantitative work on this election has begun to appear~\cite{nakamura2024,miyazaki2025,ida2025}, but little of it has shown at scale what was happening on YouTube itself.
 
Nor is it clear what kind of attention this was. Much prior work on online diffusion treats active audience response, such as sharing, as its driver~\cite{berger2012,tucker2018}. For short videos watched passively through algorithmic recommendation, however, being watched widely and being supported strongly need not coincide. Most of this work also concerns Western settings, and whether its findings apply to Japan has rarely been tested.

We therefore examine 44,331 YouTube videos relating to the nine leading candidates in the 2024 Tokyo gubernatorial election, and ask three questions.
 
\begin{itemize}
\item \textbf{RQ1}: Were clip videos associated with higher view counts, and was that association particularly strong for videos mentioning Ishimaru?
\item \textbf{RQ2}: Were high view counts accompanied by active engagement from viewers?
\item \textbf{RQ3}: Did title sentiment and visual features of the thumbnail account for the view counts of short clip videos?
\end{itemize}
 
RQ3 asks whether the view, advanced in prior work~\cite{cui2024}, that emotional appeal and visual salience account for the diffusion of video also holds for the clip format.
 
This paper makes three contributions. First, it quantitatively analyzes more than 44,000 YouTube videos on a Japanese election over four years, including 6,358 clip videos. Second, it shows that clip videos and mentions of Ishimaru were each associated with higher view counts. Third, it shows that among videos posted before the campaign, the most watched candidate ranked low on view-adjusted engagement, and that clip videos drew fewer comments relative to their views.

\section{Related Work}
The diffusion of political video has been studied mainly in terms of content and format. Regarding content, material that evokes high-arousal emotions such as anger or anxiety is shared more readily~\cite{berger2012,tucker2018}, and political posts carrying moral-emotional language spread particularly far~\cite{brady2017}. In terms of format, shorter videos tend to reach more viewers, and on YouTube, Shorts have been reported to draw far more views than regular videos~\cite{violot2024}. Political influencers have built audiences on YouTube~\cite{lewis2018}, and the business models and content strategies of such channels have themselves become an object of study~\cite{fischer2022}. How young viewers interpret political messages in YouTube videos has also been examined~\cite{bowyer2017}.

Large-scale studies of political video on YouTube have also accumulated. Some find that the recommendation algorithm leads viewers toward more radical content~\cite{ribeiro2020}, while others attribute skewed consumption to viewers' own preferences rather than to the algorithm~\cite{hosseinmardi2021}. This work has concentrated on which content is watched.

Much research treats active viewer response---likes and comments---either as the engine of diffusion or as a measure of influence. In election research, candidates with high engagement on social media have been reported to fare better at the polls~\cite{gibson2025}, although the overall relationship between social media activity and political participation is weak~\cite{boulianne2015}.

Research on consumer behavior and social media marketing distinguishes levels of user activity. The COBRAs framework, for example, classifies brand-related activities on social media into consuming, contributing and creating ~\cite{muntinga2011cobras}. In these terms, viewing a video corresponds to consuming, and liking or commenting on it to contributing.

Views and engagement need not measure the same thing. YouTube's recommender is designed to optimize predicted watch time rather than clicks or likes~\cite{covington2016}. A video surfaced by recommendation is often one the viewer did not seek out but happened upon. Furthermore, users who do not go online for news are nonetheless incidentally exposed to political information on social media~\cite{fletcher2018}. Under such passive viewing, broad reach and active engagement may diverge.

Shallow attention unaccompanied by engagement may still matter politically. People tend to favor what they have encountered repeatedly~\cite{zajonc1968}, and in elections where information about candidates is scarce, mere name recognition has been shown experimentally to raise support~\cite{kam2013}. Entertainment-oriented political content has also been reported to reach audiences with little interest in politics~\cite{baum2002}. Evaluating online attention therefore requires measuring its breadth of reach and depth of engagement as separate quantities. Few studies compare the two in the same data.

Research on clip videos has dealt mainly with revenue sharing between broadcasters and clip producers, and with producers' motives~\cite{president2022,yahoo2024kirinuki,nhk2025clipban}. Almost no existing research analyzes political clip videos at scale or examines the nature of the attention they generate.

Work on Japanese elections and social media has focused on X, formerly Twitter~\cite{miyazaki2025}. Video platforms have only recently been identified as influencing elections~\cite{nhk2025hyogo}, and quantitative analyses of political video on YouTube remain limited: the closest precedent for the present election examines discourse on the platform but not the diffusion of clip videos or the gap between reach and response~\cite{nakamura2024}. Most prior work addresses Western settings, and whether its findings carry over to Japan---where voters without a party affiliation form the largest bloc~\cite{nhk2025support}---is unclear.

This paper addresses these gaps by examining, in a single dataset, how clip videos were associated with candidate attention and whether that attention was accompanied by engagement.

\section{Research Methods}
\subsection{Data Collection}

\label{sec:data}

Data were collected through the YouTube Data API v3. To raise the quota ceiling, we applied for and were accepted into the YouTube Researcher Program.

Of the 56 candidates who ran in the 2024 Tokyo gubernatorial election, we focus on the nine who received at least 1\% of the vote. This threshold ensures our focus remains on candidates for whom online attention translated to some degree into votes. The observation window runs from the polling day of the previous gubernatorial election (July 5, 2020) to that of the 2024 election (July 7, 2024)—a span of approximately four years—in order to capture the long-run development of attention rather than the campaign period alone.

Collection proceeded in two stages. In the first stage we searched using ten terms, the nine candidate names plus “Tokyo gubernatorial election,” and retrieved the identifiers of the matching videos (late May 2025). Adding every video from the nine candidates' official channels gave 111,092 video identifiers in total, including duplicates. In the second stage we used those identifiers to retrieve metadata: view, like and comment counts, duration, title, description, tags and channel information. The second stage was carried out within a single day, June 25, 2025, so that views and engagement were measured on the same day for every video. These values are therefore cumulative up to that date, and videos posted earlier had longer to accumulate views. The regressions address this by controlling for the number of days from posting to the election.

\subsection{Preprocessing}

\begin{figure}[!t]
\centering
\includegraphics[width=\columnwidth]{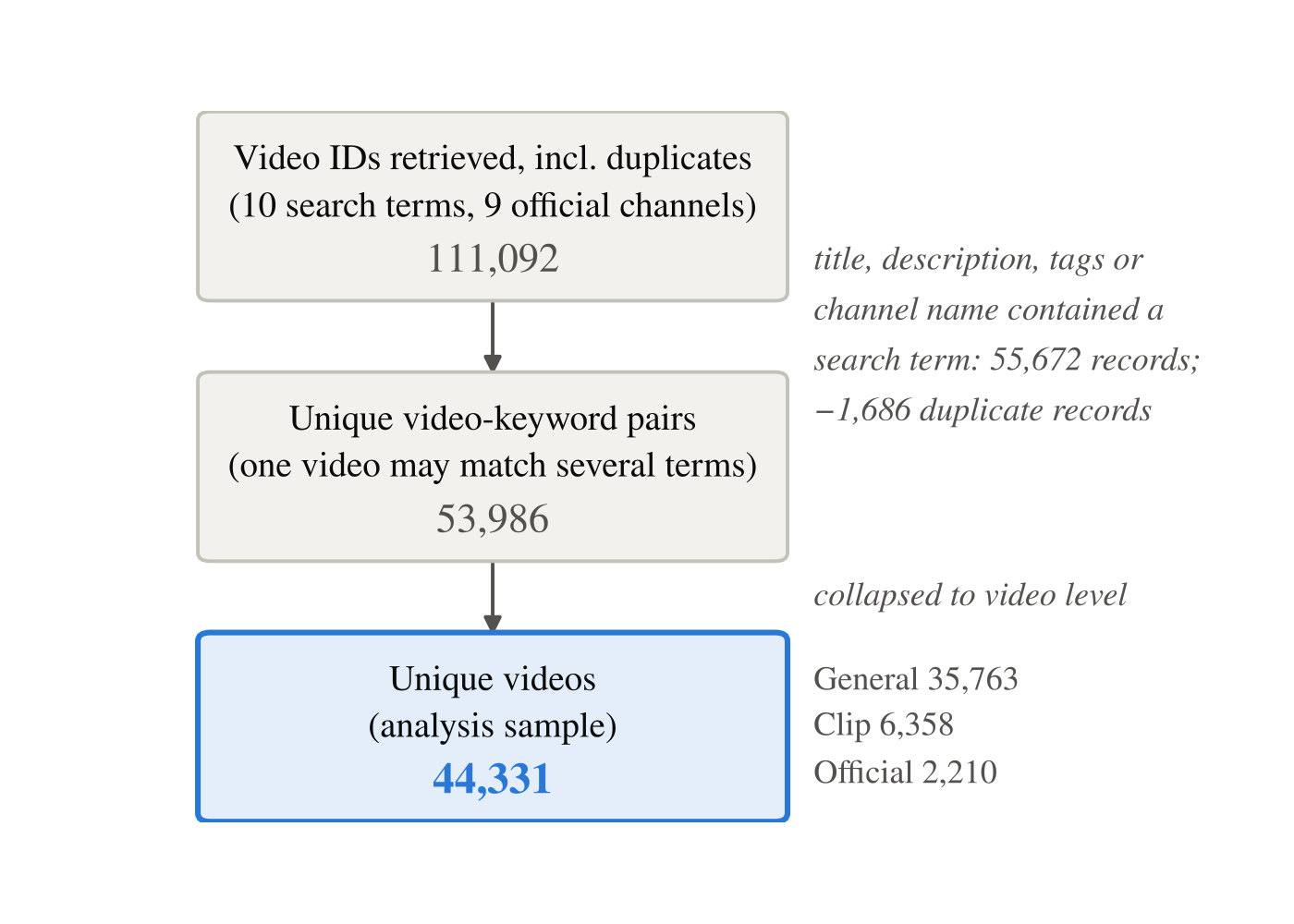}
\caption{Construction of the analysis sample.}
\label{fig:flow}
\end{figure}

Because the search API returns many loosely related videos, we kept only videos whose title, description, tags, or channel name contained a candidate name or the term “Tokyo gubernatorial election.” This left 44,331 distinct videos. A single video may mention several candidates, giving 53,986 video--keyword pairs. Of the videos, 86.3\% mention exactly one candidate and 9.5\% mention more than one. The remaining 4.2\% (1,867 videos) matched only the term “Tokyo gubernatorial election”; all candidate dummy variables are zero for these, and they form the reference group. Fig.~\ref{fig:flow} summarizes the construction of the sample.

\subsection{Classifying videos}

Videos were sorted into official, clip and general videos.

\textbf{Official videos} are posted from a candidate's own YouTube channel and mention that candidate.

\textbf{Clip videos} are defined as videos in which a third party reassembles footage already present online. To extract videos meeting this definition we treated as clip videos those that (1) contain the term \textit{kirinuki} (“clip”) in the title, description, tags, or channel name, or (2) contain both “quotation” and “YouTube”  in the title, description, or tags. The second condition reflects the convention of naming the YouTube video a clip is drawn from. The official-video condition takes precedence, and videos meeting neither condition are \textbf{general videos}, a category that includes news reports and commentary by individuals.

For this analysis, we classified videos by duration as short (60 seconds or less) or regular (longer than 60 seconds). This duration-based classification does not necessarily match YouTube’s Shorts designation.

To assess accuracy, we manually reviewed random samples of 100 clip and 100 general videos against the clip definition above, irrespective of topic. Precision was 91\% (53/58) for the keyword rule and 69\% (29/42) for the quotation rule, 82\% overall (95\% CI 73–88\%). Nine of the 100 general videos were clips (95\% CI 5–16\%), all of them unlabeled videos about Ishimaru (four) or Utsumi (five). 

\subsection{Model of view counts}

To estimate how clip videos and candidates related to view counts, we fitted an ordinary least squares regression with the logarithm of view counts as the dependent variable. No video had zero views, and all 44,331 videos enter the model.

\begin{equation}
\begin{split}
\log(\mathrm{views}_i + 1) = {} & 
\beta_0 +
\beta_1 
\mathrm{Official}_i + 
\beta_2 \mathrm{Clip}_i\\& + 
\beta_3 
\mathrm{Ishimaru}_i + 
\beta_4 (\mathrm{Clip}_i \times \mathrm{Ishimaru}_i) 
\\& + 
\boldsymbol{\gamma}' \mathbf{C}_i + \delta_1 \mathrm{Duration}_i + \delta_2 \mathrm{Days}_i + \varepsilon_i
\end{split}
\label{eq:views}
\end{equation}

$\mathrm{Official}$ and $\mathrm{Clip}$ are dummy variables for official and clip videos, with general videos as the reference category. $\mathrm{Ishimaru}$ is a dummy variable for videos mentioning Shinji Ishimaru, and $\mathbf{C}$ is a vector of dummy variables for the other eight candidates. The 1,867 videos matching only the term ``Tokyo gubernatorial election'' have all candidate dummy variables set to zero and form the reference group. Because a video may mention several candidates, the candidate dummy variables are not mutually exclusive: each coefficient contrasts videos that mention that candidate with videos that do not, holding mentions of the other candidates constant. A candidate with a similar name, Yukito Ishimaru, is included as a separate variable so that the two are not conflated. $\mathrm{Duration}$ is the length of the video in minutes and $\mathrm{Days}$ the number of days from posting to the election (July 7, 2024). The coefficient $\beta_4$ on the interaction term tests whether the association between clip videos and views differs for videos mentioning Ishimaru. 
Videos posted from the same channel share the uploader's visibility and subscriber base and are not independent of one another. Standard errors are therefore robust and clustered by channel (5,453 clusters). The largest variance inflation factor among the explanatory variables is 2.11, for the interaction term, so multicollinearity is not a concern.

\subsection{Model of engagement}

To test whether high view counts were accompanied by active engagement, we fitted the same specification with engagement as the dependent variable, adding the logarithm of view counts as a regressor.

\begin{equation}
\log(\mathrm{engagement}_i + 1) = 
\alpha_0 + 
\alpha_1 \log(\mathrm{views}_i + 1) + 
\cdots + u_i
\label{eq:engagement}
\end{equation}

Here $\mathrm{engagement}$ is either the like count or the comment count. Using the same set of independent variables as in Equation~(\ref{eq:views}) (represented by the omitted terms), we estimated the two models separately. Because a constant of 1 is added before taking logarithms, videos with zero engagement are retained. Videos whose likes are hidden or whose comments are disabled return no count from the API; in the main models we set these counts to zero so that all 44,331 videos enter every model. Excluding them instead gives slightly larger negative coefficients for Ishimaru and for clips (Table IV). Standard errors are clustered by channel as in the view-count model. Because view counts enter as a regressor, each coefficient is a difference in engagement between videos with the same number of views; a negative coefficient means that videos of that kind draw less engagement relative to their views.

\subsection{Title sentiment and thumbnail features}

An exploratory comparison carried out before this study, of the ten most and the ten least viewed short clip videos, suggested that the more viewed videos had more negative titles and more saturated thumbnails. We tested whether this holds across the full set of short clip videos ($n=2{,}631$).

Title sentiment was computed with \texttt{oseti}, a sentiment analysis library built on a Japanese polarity dictionary~\cite{ikegami2019oseti}. \texttt{oseti} returns a polarity score from $-1$ (negative) to $+1$ (positive) for each sentence; we took the mean over the sentences of a title as that video's score. For thumbnails we retrieved the 480×360 image published by YouTube (\texttt{hqdefault}), converted it to HSV color space with OpenCV and computed mean saturation. Images could not be retrieved for 202 videos, most of them deleted or made private, so the saturation analysis covers 2,429 videos.

For both measures we computed Spearman rank correlations with the logarithm of view counts and, for title sentiment, compared the top and bottom quartiles of view counts with a Mann--Whitney $U$ test.

Analyses were run in Python 3.11, with \texttt{statsmodels} for the regressions and \texttt{SciPy} for the tests.

\subsection{Robustness checks}

To examine whether the main results are sensitive to particular analytical choices, we ran six additional analyses.

\begin{enumerate}
\item \textbf{Restriction to the campaign period.} Some clip videos about Ishimaru were posted while he was mayor of Akitakata, so attention accumulated before the gubernatorial election could be driving the result. We re-estimated the view, like and comment models on videos posted during the official campaign period common to all candidates, from the official start of the campaign on June 20, 2024, to the vote on July 7 ($n=3{,}748$).
\item \textbf{Common slope in the engagement model.} The model in Equation~(\ref{eq:engagement}) assumes that the relationship between views and engagement is log-linear with a slope common to all groups. We re-estimated the like model with an interaction between $\log(\mathrm{views})$ and the Ishimaru dummy variable.
\item \textbf{Regular videos only.} Because views are counted on a different basis for Shorts, which during our observation period were limited to 60 seconds, we re-estimated the view, like and comment models on videos longer than 60 seconds ($n=34{,}377$).
\item \textbf{Period-specific estimates.} We re-estimated all three models separately for videos posted before 2024 ($n=31{,}802$), between January 1 and June 19, 2024 ($n=8{,}781$), and during the campaign period.
\item \textbf{Alternative clip definitions.} We re-estimated all models with clip videos defined by the \textit{kirinuki}-keyword rule alone, and with general videos restricted to channels that posted no detected clip ($n=37{,}477$).
\item \textbf{Hidden likes and disabled comments.} We re-estimated the like and comment models excluding videos whose likes were hidden ($n=43{,}211$) or whose comments were disabled ($n=43{,}098$), respectively.
\end{enumerate}

\section{Results}
\label{sec:results}

\subsection{Concentration of attention across candidates}

\begin{figure}[!t]
\centering
\includegraphics[width=\columnwidth]{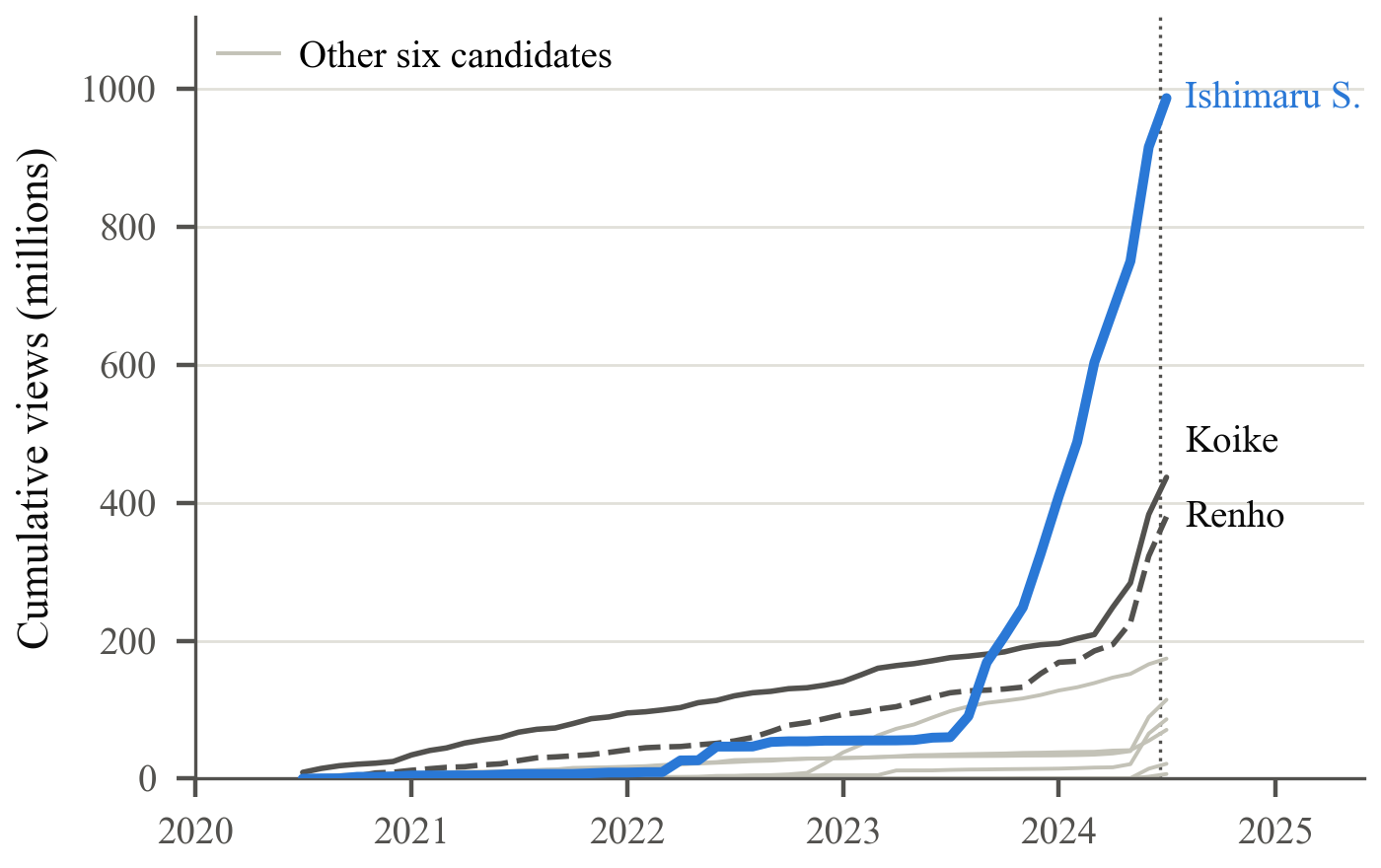}
\caption{Cumulative views by candidate. Values are cumulative over videos posted
up to each month, measured on June 25, 2025. The dotted line marks the start of
the official campaign period (June 20, 2024).}
\label{fig:cumulative}
\end{figure}

Cumulative views were heavily concentrated on one candidate. Videos mentioning
Ishimaru accumulated about 990 million views, 2.3 times the total for Koike
(about 440 million), the second-ranked candidate (Fig.~\ref{fig:cumulative}). The
ordering by number of videos differed: Koike had the most related videos (15,335)
and Ishimaru less than half that number (6,685). The concentration arose in views
rather than in the volume of uploads.

\begin{figure}[!t]
\centering
\includegraphics[width=\columnwidth]{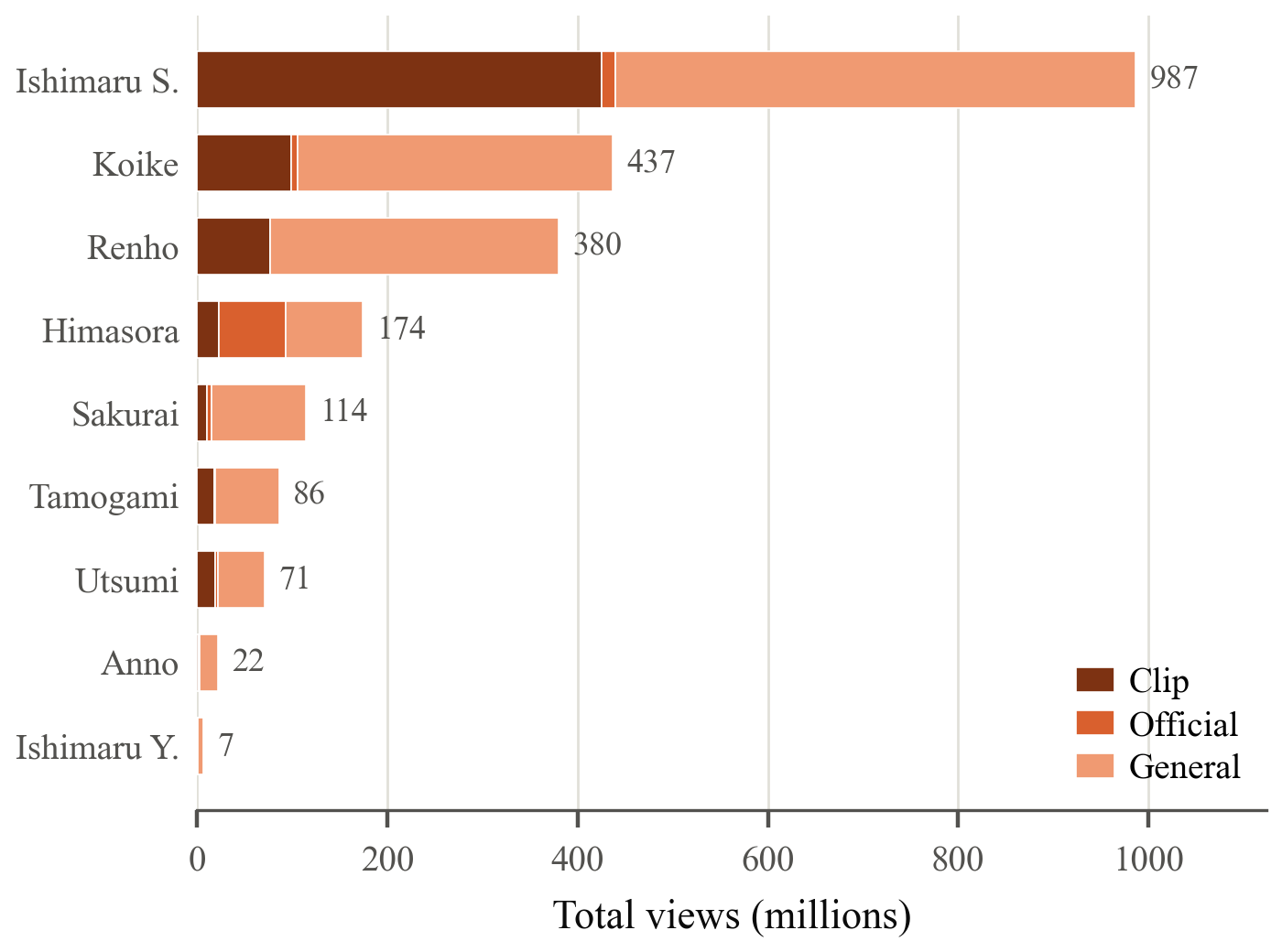}
\caption{Total views by candidate and video type. Videos mentioning several
candidates are counted once for each, so the bars are not mutually exclusive and
do not sum to the dataset total.}
\label{fig:bytype}
\end{figure}

Clip videos accounted for much of this concentration. Clip videos mentioning Ishimaru alone
accumulated about 425 million views, comparable to the total for all videos
mentioning Koike (Fig.~\ref{fig:bytype}). At the channel level, 17 of the 30
channels with the highest cumulative views posted clip videos, and 10 of these
drew more than half their total views from clips. Four of those channels included
``Ishimaru'' in their channel names, indicating the emergence of third-party uploaders specializing in a
single candidate.

\subsection{Clip videos, candidates and view counts}

Table~\ref{tab:counts} reports the number of videos by type. Clip videos (6,358)
are about 2.9 times as numerous as official videos (2,210), and 41.4\% of them
are 60 seconds or less---a proportion markedly higher than for general
(19.7\%) or official (12.3\%) videos.

\begin{table}[!t]
\caption{Number of Videos by Type}
\label{tab:counts}
\centering
\begin{tabular}{lrrrr}
\toprule
Type & Short & Regular & Total & Short (\%) \\
\midrule
General  & 7,051 & 28,712 & 35,763 & 19.7 \\
Clip     & 2,631 &  3,727 &  6,358 & 41.4 \\
Official &   272 &  1,938 &  2,210 & 12.3 \\
\midrule
Total    & 9,954 & 34,377 & 44,331 & 22.5 \\
\bottomrule
\end{tabular}
\end{table}

To control simultaneously for video format, duration, timing, and the candidates mentioned, we regressed the logarithm of view counts on these variables (Table~\ref{tab:views}).

In the full-period model ($n=44{,}331$), holding the other variables constant, clip videos not mentioning Ishimaru received about 1.8 times as many views as general videos ($\beta=0.577$, $p=.020$), and non-clip videos mentioning Ishimaru received about 2.6 times as many views as those that did not mention him ($\beta=0.974$, $p=.030$). The interaction between clip videos and mentions
of Ishimaru was not significant ($\beta=0.116$, $p=.854$): we find no evidence that the association between the clip format and views was especially strong for videos mentioning him.

\begin{table}[!t]
\caption{Regression on Log View Counts}
\label{tab:views}
\centering
\begin{tabular}{lrrrr}
\toprule
& \multicolumn{2}{c}{Full period} & \multicolumn{2}{c}{Campaign period} \\
\cmidrule(lr){2-3} \cmidrule(lr){4-5}
Variable & $\beta$ & $p$ & $\beta$ & $p$ \\
\midrule
Official video         &  1.974 & $<$.001 &  1.006 & .001 \\
Clip video             &  0.577 & .020    &  1.329 & .045 \\
Ishimaru S.            &  0.974 & .030    &  1.491 & $<$.001 \\
Clip $\times$ Ishimaru &  0.116 & .854    & $-$0.657 & .390 \\
Duration (min)         &  0.008 & $<$.001 &  0.015 & $<$.001 \\
Days to election       & $-$0.0007 & .005 &  0.028 & .034 \\
Koike                  & $-$0.359 & .192  & $-$0.009 & .961 \\
Renho                  &  0.629 & .031    &  0.020 & .916 \\
Himasora               &  0.504 & .069    & $-$0.977 & $<$.001 \\
Sakurai                & $-$1.415 & $<$.001 & 0.084 & .835 \\
Tamogami               & $-$0.148 & .545  & $-$0.508 & .015 \\
Utsumi                 & $-$0.313 & .482  & $-$0.452 & .316 \\
Anno                   &  0.170 & .743    & $-$0.322 & .207 \\
Ishimaru Y.            & $-$1.602 & .025  &  0.192 & .677 \\
\midrule
$N$                    & \multicolumn{2}{c}{44,331} & \multicolumn{2}{c}{3,748} \\
$R^2$                  & \multicolumn{2}{c}{0.166}  & \multicolumn{2}{c}{0.134} \\
\bottomrule
\end{tabular}
\par\vspace{2pt}
\begin{minipage}{\columnwidth}
\footnotesize Dependent variable: $\log(\mathrm{views}+1)$. For dummy variables, $\exp(\beta)$ is the ratio of views; for example, $\exp(0.974)\approx 2.6$. Reference categories are general videos for video type and, for each candidate, videos that do not mention that candidate. Because the model includes the Clip $\times$ Ishimaru term, Clip video compares clip with general videos among videos not mentioning Ishimaru, and Ishimaru S.\ compares videos that do and do not mention him among non-clip videos. Among clip videos, the Ishimaru difference is the sum of the Ishimaru S.\ and Clip $\times$ Ishimaru coefficients. Standard errors are robust and clustered by channel (5,453 clusters for the full period, 1,069 for the campaign period). The intercept is omitted.
\end{minipage}
\end{table}

\textbf{Robustness.} 
To remove the influence of videos posted during his tenure as mayor, we restricted the sample to the official campaign period ($n=3{,}748$). The coefficients for clip videos ($\beta=1.329$, $p=.045$) and for Ishimaru ($\beta=1.491$, $p<.001$) both remained significant, and the interaction remained non-significant. Because these videos also accumulated views for about a year after the election, this restriction does not isolate attention during the campaign. 

The association between clip videos and views remained when general videos were restricted to channels that posted no detected clip ($\beta=0.742$, $p=.011$). Among videos longer than 60 seconds, the coefficient remained positive but was not statistically significant ($\beta=0.554$, $p=.052$). The association was not statistically significant when clips were defined by the keyword rule alone ($\beta=0.316$, $p=.306$; Table~\ref{tab:robustness}).

\subsection{Divergence between views and engagement}

To test whether high view counts were accompanied by active engagement, we
regressed the logarithms of like and comment counts on the logarithm of
view counts and the same covariates (Table~\ref{tab:engagement}), using all
44,331 videos.

The coefficient on views is 0.829 in the like model and 0.618 in the comment
model, both below one: as views rise, likes and comments per view both decline.

\begin{table}[!t]
\caption{Regression on Log Likes and Log Comments, Controlling for Views}
\label{tab:engagement}
\centering
\begin{tabular}{lrrrr}
\toprule
& \multicolumn{2}{c}{Likes} & \multicolumn{2}{c}{Comments} \\
\cmidrule(lr){2-3} \cmidrule(lr){4-5}
Variable & $\beta$ & $p$ & $\beta$ & $p$ \\
\midrule
Log views              &  0.829 & $<$.001 &  0.618 & $<$.001 \\
Official video         &  0.307 & .199    & $-$0.344 & .117 \\
Clip video             & $-$0.133 & .155  & $-$0.306 & .003 \\
Ishimaru S.            & $-$0.440 & $<$.001 & $-$0.524 & $<$.001 \\
Clip $\times$ Ishimaru &  0.031 & .814    &  0.074 & .706 \\
Duration (min)         & $-$0.0004 & .577  & $-$0.0025 & .002 \\
Days to election       & $-$0.0002 & .003 &  0.0000 & .552 \\
Koike                  & $-$0.243 & .007  &  0.065 & .474 \\
Renho                  & $-$0.050 & .654    &  0.288 & .004 \\
Himasora               &  0.482 & $<$.001 &  0.414 & $<$.001 \\
Sakurai                &  0.629 & $<$.001 &  0.338 & .002 \\
Tamogami               &  0.255 & $<$.001 &  0.163 & .071 \\
Utsumi                 &  0.472 & $<$.001 & $-$0.121 & .541 \\
Anno                   & $-$0.337 & .023  & $-$0.267 & .048 \\
Ishimaru Y.            & $-$1.375 & $<$.001 & $-$0.417 & .034 \\
\midrule
$N$                    & \multicolumn{2}{c}{44,331} & \multicolumn{2}{c}{44,331} \\
$R^2$                  & \multicolumn{2}{c}{0.822}  & \multicolumn{2}{c}{0.696} \\
\bottomrule
\end{tabular}
\par\vspace{2pt}
\begin{minipage}{\columnwidth}
\footnotesize Dependent variables: $\log(\mathrm{likes}+1)$ and $\log(\mathrm{comments}+1)$. Because log views is included, each coefficient compares videos with the same number of views. For dummy variables, $\exp(\beta)$ is the ratio of likes or comments; for example, $\exp(-0.440)\approx 0.64$, or about 36\% fewer likes. Reference categories and the reading of the Clip video and Ishimaru S.\ coefficients are as in Table~\ref{tab:views}. Standard errors are robust and clustered by channel (5,453 clusters). The intercept is omitted.
\end{minipage}
\end{table}

\begin{table*}[!t]
\centering
\caption{Robustness of the Main Coefficients}
\label{tab:robustness}
\footnotesize
\setlength{\tabcolsep}{5pt}
\begin{tabular}{lrrrrrrr}
\hline
 & & \multicolumn{2}{c}{Views} & \multicolumn{2}{c}{Likes} & \multicolumn{2}{c}{Comments} \\
Specification & $N$ & Clip & Ishimaru S. & Clip & Ishimaru S. & Clip & Ishimaru S. \\
\hline
Main model (Tables II, III) & 44,331 & $0.577^{*}$ & $0.974^{*}$ & $-0.133$ & $-0.440^{***}$ & $-0.306^{**}$ & $-0.524^{***}$ \\
Excluding hidden likes / disabled comments & 43,211 / 43,098 & --- & --- & $-0.193^{*}$ & $-0.500^{***}$ & $-0.390^{***}$ & $-0.585^{***}$ \\
Clips: keyword rule only & 44,331 & $0.316$ & $1.020^{*}$ & $-0.340^{**}$ & $-0.477^{***}$ & $-0.514^{***}$ & $-0.598^{***}$ \\
General: channels with no detected clip & 37,477 & $0.742^{*}$ & $0.785$ & $-0.082$ & $-0.413^{***}$ & $-0.329^{**}$ & $-0.510^{***}$ \\
Regular videos only & 34,377 & $0.554^{\dagger}$ & $0.985^{***}$ & $-0.118$ & $-0.517^{***}$ & $-0.244^{*}$ & $-0.172$ \\
Posted before 2024 & 31,802 & $0.328$ & $0.183$ & $-0.211^{\dagger}$ & $-0.644^{**}$ & $-0.401^{**}$ & $-0.812^{***}$ \\
Posted Jan 1 -- Jun 19, 2024 & 8,781 & $1.310^{***}$ & $1.173^{***}$ & $0.003$ & $-0.488^{***}$ & $0.011$ & $-0.561^{***}$ \\
Campaign period (Jun 20 -- Jul 7, 2024) & 3,748 & $1.329^{*}$ & $1.491^{***}$ & $0.265$ & $0.451^{*}$ & $-0.116$ & $0.150$ \\
\hline
\end{tabular}

\vspace{2pt}
\parbox{\textwidth}{\footnotesize Coefficients from the specifications of Tables II and III; the likes and comments models include log views. In the second row, hidden likes and disabled comments are excluded from the likes and comments models, respectively; the view model is unaffected. Standard errors are clustered by channel. $^{\dagger}p<.10$, $^{*}p<.05$, $^{**}p<.01$, $^{***}p<.001$.}
\end{table*}

\begin{figure}[!t]
\centering
\includegraphics[width=\columnwidth]{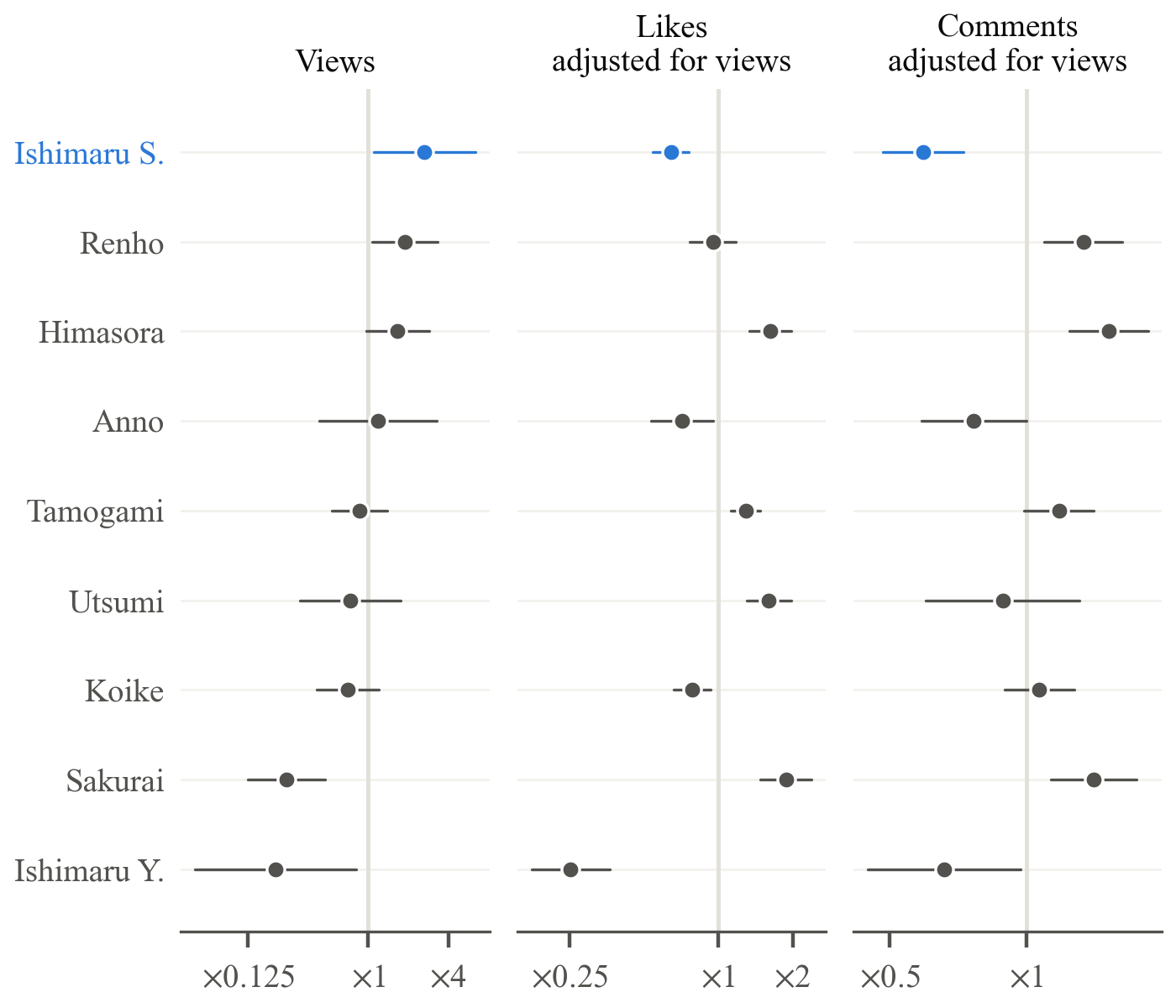}
\caption{Coefficients by candidate. From left: views, likes adjusted for views,
and comments adjusted for views. Estimates are shown as ratios, $\exp(\beta)$,
on a log scale. Points are point estimates and lines show 95\% confidence
intervals. The reference group for each candidate is videos that do not
mention that candidate.}
\label{fig:coef}
\end{figure}

Holding views and the other conditions constant, non-clip videos mentioning
Ishimaru received about 36\% fewer likes ($\beta=-0.440$, $p<.001$) and about
41\% fewer comments ($\beta=-0.524$, $p<.001$) than those that did not mention
him. Sakurai shows the opposite pattern. Holding the other conditions constant,
videos mentioning him received about a quarter as many views as those that did
not (Table~\ref{tab:views}, $\beta=-1.415$), yet, holding views as well, about
1.9 times as many likes ($\beta=0.629$, $p<.001$) and about 1.4 times as many
comments ($\beta=0.338$, $p=.002$). Himasora is high on both measures (about 1.6
times for likes and 1.5 times for comments), while Utsumi is high on likes but shows no
significant difference on comments ($\beta=-0.121$, $p=.541$). The candidates watched most
widely were not those engaged with most actively (Fig.~\ref{fig:coef}).

Holding views and the other conditions constant, clip videos not mentioning
Ishimaru received about 26\% fewer comments than general videos
($\beta=-0.306$, $p=.003$); their coefficient for likes is also negative but not
significant ($\beta=-0.133$, $p=.155$). The divergence between views and
engagement is largest at the level of the candidate, but for comments it extends
to the video format as well. Official videos did not differ significantly from
general videos on either measure.

\textbf{Robustness.}
The interaction between $\log(\mathrm{views})$ and the Ishimaru dummy variable in the like model was positive and significant ($\beta=0.067$, $p=.032$): the shortfall in likes is somewhat smaller for videos with more views, so the common-slope model reports an average gap rather than a constant one. Because views are counted on a different basis for Shorts, we also re-estimated the model on regular videos alone ($n=34{,}377$): the coefficient for Ishimaru is $-0.517$ ($0.596$ times, $p<.001$), slightly larger than in the full sample. 

The shortfall in likes is therefore unlikely to be  an artifact of how Shorts views are counted. In the same analysis, clip videos not mentioning Ishimaru received fewer view-adjusted comments than general videos ($\beta=-0.244$, $p=.040$), while the difference in likes was not significant ($\beta=-0.118$, $p=.222$). The shortfall in likes for videos mentioning Ishimaru was also robust to both alternative clip definitions and to the exclusion of hidden likes ($\beta$ from $-0.413$ to $-0.500$, all $p<.001$; Table~\ref{tab:robustness}). It was present among videos posted before 2024 and between January 1 and June 19, 2024, but reversed among campaign-period videos ($\beta=0.451$, $p=.013$). As with the view-count model, these counts include about a year of post-election engagement. The shortfall in comments was likewise robust except among regular videos, where it was smaller and not significant ($\beta=-0.172$, $p=.157$), and during the campaign ($\beta=0.150$, $p=.269$).

\subsection{Title sentiment and thumbnail features}

The relationship seen in the exploratory comparison did not replicate across the
full set. Title sentiment was not significantly correlated with view counts
(Spearman $\rho=-0.029$, $p=.142$), and the top and bottom quartiles did not
differ ($p=.167$). Thumbnail saturation was likewise unrelated to view counts
($\rho=-0.006$, $p=.765$, $n=2{,}429$). Within the range of dictionary-based
sentiment scores and thumbnail saturation, then, we found no general
association that accounts for the view counts of short clip videos. This stands in
contrast to commercial video, where the sentiment of titles and thumbnails has
been shown to predict views~\cite{cui2024}. Titles of political
clip videos often use irony and platform-specific idioms, which a polarity dictionary may not capture; the null result therefore does not show that sentiment is irrelevant, and context-dependent classification may be needed. The difference observed in the exploratory comparison may be confined to a small number of
outlying videos.

\section{Discussion}
\label{sec:discussion}

\subsection{Broad but shallow attention}

The central result is that the most watched candidate and the most actively
engaged-with candidate were not the same. Among non-clip videos, those mentioning Ishimaru received about 2.6 times as many views with other conditions held constant, yet about 36\% fewer view-adjusted likes than those that did not. Videos mentioning Sakurai or Himasora, by
contrast, received more view-adjusted likes and comments.

This contrast suggests at least two patterns in the metrics. One combines many views with little engagement per view, which we call broad but shallow; the other combines fewer views with more engagement per view, which we call narrow but deep. Views count viewings rather than distinct viewers, and likes and comments do not measure the strength of political support, so these terms describe the metrics, not the audience. The broad but shallow pattern held for videos posted before the campaign; among campaign-period videos, those mentioning Ishimaru received more likes per view. Because all counts were measured about a year after the election, this reversal cannot be attributed to engagement during the campaign itself.

Given that YouTube's recommender is built around watch time~\cite{covington2016} and that incidental exposure is widespread on social media~\cite{fletcher2018}, Ishimaru's videos may have been watched by viewers who
happened upon them rather than supporters who sought them out. Which route
carried them cannot be identified from our data: presentation by the
recommendation algorithm, sharing on other platforms, and search traffic driven by
newsworthiness are all consistent with the same observation. Traffic sources are
not available through the platform's public API and can be seen only by the
channel owner.

This reading may resolve the apparent tension between low engagement and
Ishimaru's 1.66 million votes. Where information about candidates is scarce, mere
name recognition has been shown to raise support~\cite{kam2013}. For a candidate
previously unknown nationally, broad exposure on YouTube may have served to make
his name and persona widely known, whether or not it produced committed supporters. We do
not measure the relationship between viewing and voting, so this remains a
hypothesis for future work.

\subsection{Clip videos and attention to Ishimaru}

The interaction between clip videos and mentions of Ishimaru was not significant
in the main models or in the campaign-period models. From this result we cannot conclude that the clip format was
associated especially strongly with Ishimaru; nor does it show that the strength
of the association was the same. Understanding the prominence of his cumulative
views requires considering, beyond the association with clip videos, the attention
to Ishimaru himself and the number of clip videos about him (2,321, the most
for any candidate).

Four clip channels including “Ishimaru” in their names were among the channels with the highest cumulative views (Section IV-A). If producers are motivated by revenue, they would select candidates likely to draw
views and produce clip videos of them in volume, so a cycle may have arisen in
which supply concentrates further on candidates already receiving attention. Whether attention drew supply or supply generated attention cannot be determined from our data.

In this design, redistribution by third parties appears on the explanatory side
rather than as an engagement measure, because a clip video is itself the product
of a third party reassembling and reposting existing footage. That videos
mentioning Ishimaru were clipped in such volume, to a cumulative total comparable
to all videos mentioning Koike, is itself an indication of the scale of that
redistribution.

\subsection{The Japanese context and generalizability}

The association between negative titles and view counts observed in the
exploratory analysis was not confirmed across all collected clip videos of 60
seconds or less. Comparison with the Western finding that high-arousal emotion
promotes diffusion~\cite{berger2012} must, however, allow for the difference in
measures: the positive--negative polarity of titles measured here is distinct
from emotional arousal, so this result alone does not show that the finding
fails to hold in Japan. Whether characteristics of Japanese viewers play a role
should be examined through comparison with other countries, after validating the
measurement accuracy of dictionary-based sentiment analysis. The divergence
between views and engagement, by contrast, need not be specific to Japan.
Because third parties reassemble and repost existing footage in other regions and on other platforms as well, as in the 2024 U.S. presidential election~\cite{stelter2024}, testing whether a similar divergence appears there is a task for future work.

\subsection{Limitations}

This study has several limitations. First, it is an analysis of observational
data and does not establish that clip videos caused higher view counts. Second,
we do not measure the relationship between viewing and actual voting behavior.
Third, the analysis covers a single platform and a single election. Fourth, views
and engagement are cumulative values measured about a year after the election and
include views accumulated afterward. Although we control for the number of days to the election and re-estimate the models on campaign-period videos, even those videos accumulated about a year of post-election views and engagement before measurement, so no result here, including those for the campaign period, describes activity during the campaign alone.
Fifth, engagement metrics are limited to likes and comments: direct
measures of redistribution such as shares and repeat views are not available
through the platform's public API and can be seen only by the channel owner.
Furthermore, we do not analyze comment content. Sixth, title sentiment is measured by a dictionary-based method that may not capture irony or context-dependent expressions.
Seventh, the keyword rules miss some clips: about 9\% of sampled general videos were clips (95\% CI 5--16\%), all of them about Ishimaru or Utsumi. Clip channels operated or commissioned by a campaign also cannot be identified from public metadata. Eighth, engagement per view may differ between channels, for example between news outlets and individual commentators. Our models account for videos from the same channel only through clustered standard errors; channel fixed effects or channel-level controls are left for future work.
Future work includes applying the same approach to other elections in which clip
videos have been held to matter, such as the 2024 Hyogo gubernatorial election;
analyzing comment content; comparing platforms such as TikTok; and linking
viewing exposure to voting behavior through survey data.

\section{Conclusion}

We analyzed 44,331 YouTube videos related to the nine leading candidates in the
2024 Tokyo gubernatorial election and examined how clip videos were associated with
the attention candidates received. Clip videos and mentions of Shinji Ishimaru were
each associated with higher view counts. Their interaction was not significant,
providing no evidence that the association between clip videos and views was
especially strong for videos mentioning Ishimaru. Among non-clip videos, those mentioning Ishimaru received about 36\% fewer view-adjusted likes and 41\% fewer view-adjusted comments than those that did not, a shortfall concentrated in videos posted before the campaign.

These results show that the scale of views of videos about a candidate and the
amount of view-adjusted engagement need not coincide. Understanding political
influence online requires distinguishing how many views content received from how much engagement it drew.

\section*{Statements}
\textbf{Data Availability}: The collected dataset and code are available from the authors upon request.

\textbf{Ethical Statement}: All data were collected through the public YouTube Data API and consist only of publicly available video metadata, in accordance with YouTube's Terms of Service.

\textbf{Disclosure Statement}: The authors declare no potential conflicts of interest with respect to the research, authorship, and/or publication of this article.


\end{document}